\documentclass[superscriptaddress,reprint,amsmath,amssymb,showkeys]{revtex4-2}
\setcitestyle{super}
\usepackage{graphicx}
\usepackage{dcolumn}
\usepackage{bm}
\usepackage{color}
\usepackage{float}
\usepackage[utf8]{inputenc}
\usepackage[mathlines]{lineno}
\usepackage{multirow}
\usepackage{amsmath}
\usepackage{hyperref}
\usepackage{soul,xcolor}
\usepackage{makecell}

\soulregister\cite7
\soulregister\ref7
\setstcolor{blue}
\newcommand{\hmn}[1]{
 \ensuremath{\begingroup\setupHMN #1\endgroup}%
}

\newcommand{\setupHMN}{%
 \doHMN{-}{\HMNoverline}%
 \doHMN{*}{\HMNminverse}%
 \doHMN{i}{\infty}
}

\newcommand{\doHMN}[2]{%
 \begingroup\lccode`~=`#1
 \lowercase{\endgroup\let~}#2%
 \mathcode`#1="8000
}

\newcommand{\HMNminverse}[1]{\frac{#1}{m}}
\newcommand{\HMNoverline}[1]{\mkern1mu\overline{\mkern-1mu#1\mkern-1mu}\mkern1mu}

\makeatletter
\newcommand*{\addFileDependency}[1]{
 \typeout{(#1)}
 \@addtofilelist{#1}
 \IfFileExists{#1}{}{\typeout{No file #1.}}
}
\makeatother

\newcommand{\wn}{cm$^{-1}$}
\newcommand{\wns}{cm$^{-1}$ }

\begin{document}

\title{Anharmonic Lattice Dynamics in Sodium Ion Conductors}

\author{Thomas M. Brenner}
\affiliation{Department of Chemical and Biological Physics, Weizmann Institute of Science, Rehovot 76100, Israel}
\author{Manuel Grumet}
\affiliation{Department of Physics, Technical University of Munich, 85748 Garching, Germany}
\author{Paul Till}
\affiliation{Institute for Inorganic and Analytical Chemistry, University of Muenster, Münster 48149, Germany}
\author{Maor Asher}
\affiliation{Department of Chemical and Biological Physics, Weizmann Institute of Science, Rehovot 76100, Israel}
\author{Wolfgang G. Zeier}
\affiliation{Institute for Inorganic and Analytical Chemistry, University of Muenster, Münster 48149, Germany}
\author{David A. Egger}
\affiliation{Department of Physics, Technical University of Munich, 85748 Garching, Germany}
\author{Omer Yaffe}
\affiliation{Department of Chemical and Biological Physics, Weizmann Institute of Science, Rehovot 76100, Israel}


\begin{abstract}

We employ THz-range temperature-dependent Raman spectroscopy and first-principles lattice-dynamical calculations to show that the undoped sodium ion conductors Na$_3$PS$_4$ and isostructural Na$_3$PSe$_4$ both exhibit anharmonic lattice dynamics.
The anharmonic effects in the compounds involve coupled host lattice – Na$^+$ ion dynamics that drive the tetragonal-to-cubic phase transition in both cases, but with a qualitative difference in the anharmonic character of the transition.
Na$_3$PSe$_4$ shows almost purely displacive character with the soft modes disappearing in the cubic phase as the change of symmetry shifts these modes to the Raman-inactive Brillouin zone boundary.
Na$_3$PS$_4$ instead shows order-disorder character in the cubic phase, with the soft modes persisting through the phase transition and remaining active in Raman in the cubic phase, violating Raman selection rules for that phase. 
Our findings highlight the important role of coupled host lattice – mobile ion dynamics in vibrational instabilities that are coincident with the exceptional conductivity in these Na$^+$ ion conductors.

\end{abstract}

\date{\today}

\maketitle

Solid-state ion conductors (SSICs) show great promise for enabling next generation energy storage devices that are safer and more energy dense~\cite{Manthiram:2017aa}. 
The development of new, stable, and highly conductive SSICs requires a clear understanding of which material properties are essential to ion conductivity.
Intensive research in recent years indicates that many highly conductive SSIC materials exhibit lattice dynamical phenomena consistent with strong anharmonicity~\cite{Smith:2020aa,Zhang:2019aa,Zhang:2020aa,Famprikis:2019a,Brenner:2020a,Gupta:2021a,Niedziela:2019aa,Ding:2020aa,Kweon:2017aa,Adelstein:2016aa,Duchene:2019aa}.
Anharmonicity refers to the coupling that occurs between vibrational normal modes (or, phonons) of the lattice~\cite{dove2003structure,Califano1981}.
Many materials exhibit mild anharmonicity which is expressed in thermal expansion, thermal conductivity, and finite phonon lifetimes.
In contrast, SSICs are expected to exhibit strong anharmonicity because the process of hopping takes the mobile ion into a strongly anharmonic region of its potential energy, where it may couple to other vibrations present in the crystal~\cite{Rice:1958aa,Vineyard:1957aa,Salamon:1979aa,Brenner:2020a}.

The strongly anharmonic behavior of SSIC materials was shown to have different expressions.
For instance, plastic crystal phases and corresponding paddle wheel effects~\cite{Jansen:1991aa,Lunden:1988aa} have been proposed in highly conductive, ionically bonded sulfide electrolytes~\cite{Smith:2020aa,Zhang:2019aa,Zhang:2020aa,Famprikis:2019a} and hydroborates~\cite{Kweon:2017aa}, and relaxation phenomena tied to anharmonic effects have been observed in soft host lattices~\cite{Brenner:2020a,Zhang:2019aa,Zhang:2020aa}.
In light of the latter, in a previous work on the structural dynamics of $\alpha$-AgI (an archetypal SSIC) we proposed that host-lattice anharmonicity should be used as an experimental indicator for the design of new superionic conductors~\cite{Brenner:2020a}.

In a recent important study, Gupta \textit{et al.} used neutron scattering and molecular dynamics (MD) to establish the connection between anharmonic phonon dynamics and superionic conductivity in Na$_3$PS$_4$~\cite{Gupta:2021a}.
This material is a parent compound of Na$_{3-x}$P$_{1-x}$W$_x$S$_4$, a record Na$^+$ conductivity compound in which Na$^+$ vacancies have been introduced through tungsten doping~\cite{Fuchs:2020a,Hayashi:2019vr}.
They identified soft-modes that stabilize the cubic phase. 
Furthermore, they demonstrated how these strongly anharmonic modes enable Na$^+$-ions to hop along the minimum energy pathways.

However, neutron scattering is a costly experimental method that has many technical constraints~\cite{doi:10.1063/1.3680104}.
To implement anharmonicity as an indicator in material design, it is important to establish more accessible experimental characterization tools.
To that end, Raman scattering spectroscopy is a very promising table-top technique that  benefits from the ease of manipulating and detecting visible light with modern optics and microscopy.
Therefore, it is useful even for very small sample size/weight and can be measured at a wide range of temperatures and pressures.
As such, it is ideal for large throughput and spatial mapping with diffraction limited resolution.   
Importantly, in recent years we demonstrated that THz-range Raman scattering combined with first-principles computations is very effective in elucidating the atomic-scale mechanisms that lead to anharmonic motion in solids~\cite{Yaffe:2017aa,Brenner:2020a,Asher2020,Sharma:2020a,Sharma:2020b}.  

\begin{figure}
     \centering
     \includegraphics[width=0.85\linewidth]{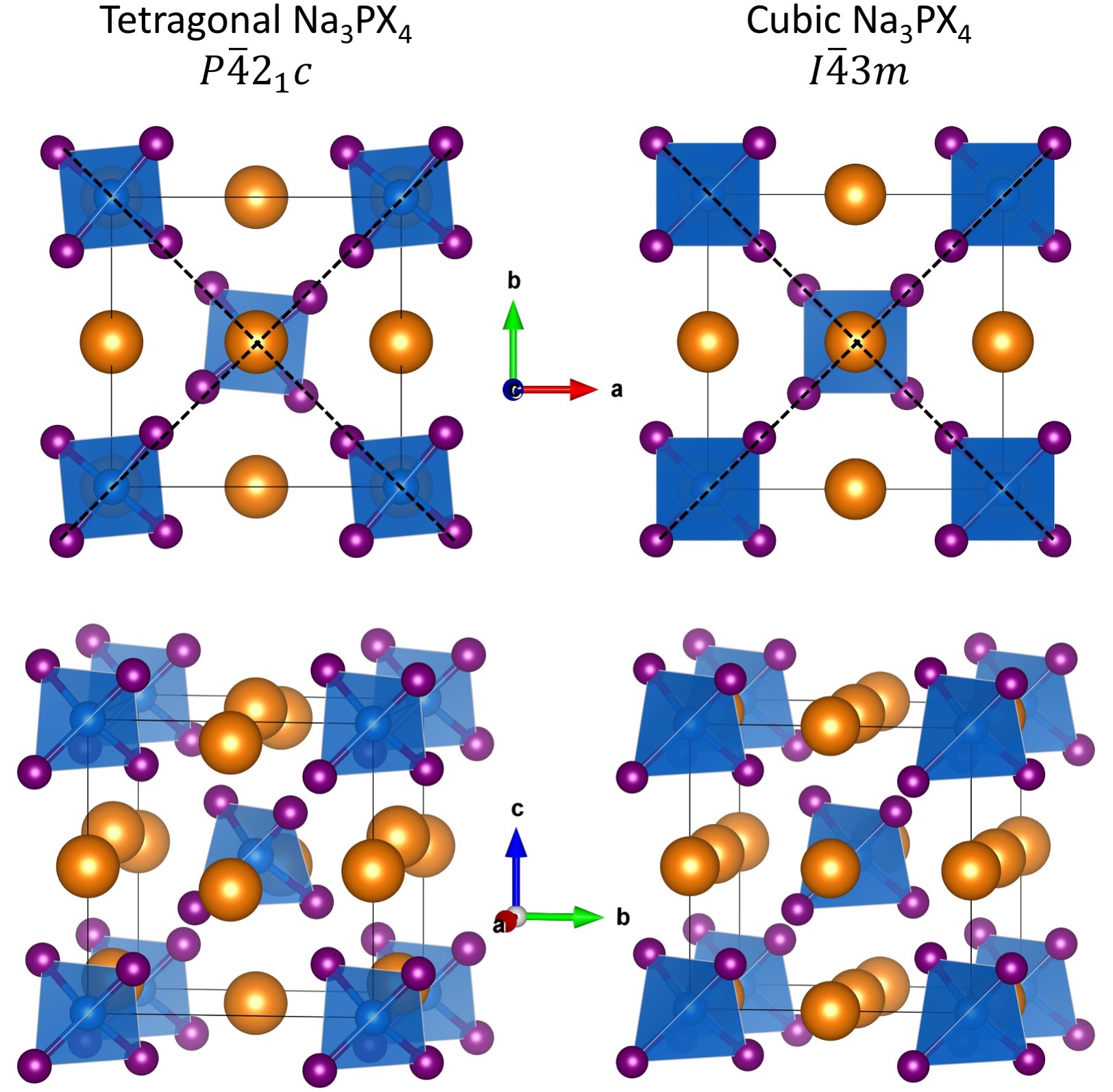}
     \caption{Schematic representation of the tetragonal~\cite{Nishimura:2017a} and cubic~\cite{Bo:2016tj} structures of the Na$_3$PCh$_4$ (Ch=S, Se) crystal system. Na$^+$ ions are shown in orange, P atoms in blue (and within blue-shaded PCh$_{4}^{3-}$ tetrahedra), and Ch atoms in purple color. The black-dotted lines indicate the presence (t-phase) or absence (c-phase) of tetrahedral tilting about the c-axis. The offset of Na$^+$ ions along the c-axis in the tetragonal compared to the cubic phase can be seen in the bottom panel.}
     \label{Fig:Na3PCh4_Structure}
 \end{figure}

\begin{figure*}
     \centering
     \includegraphics[width=0.9\textwidth]{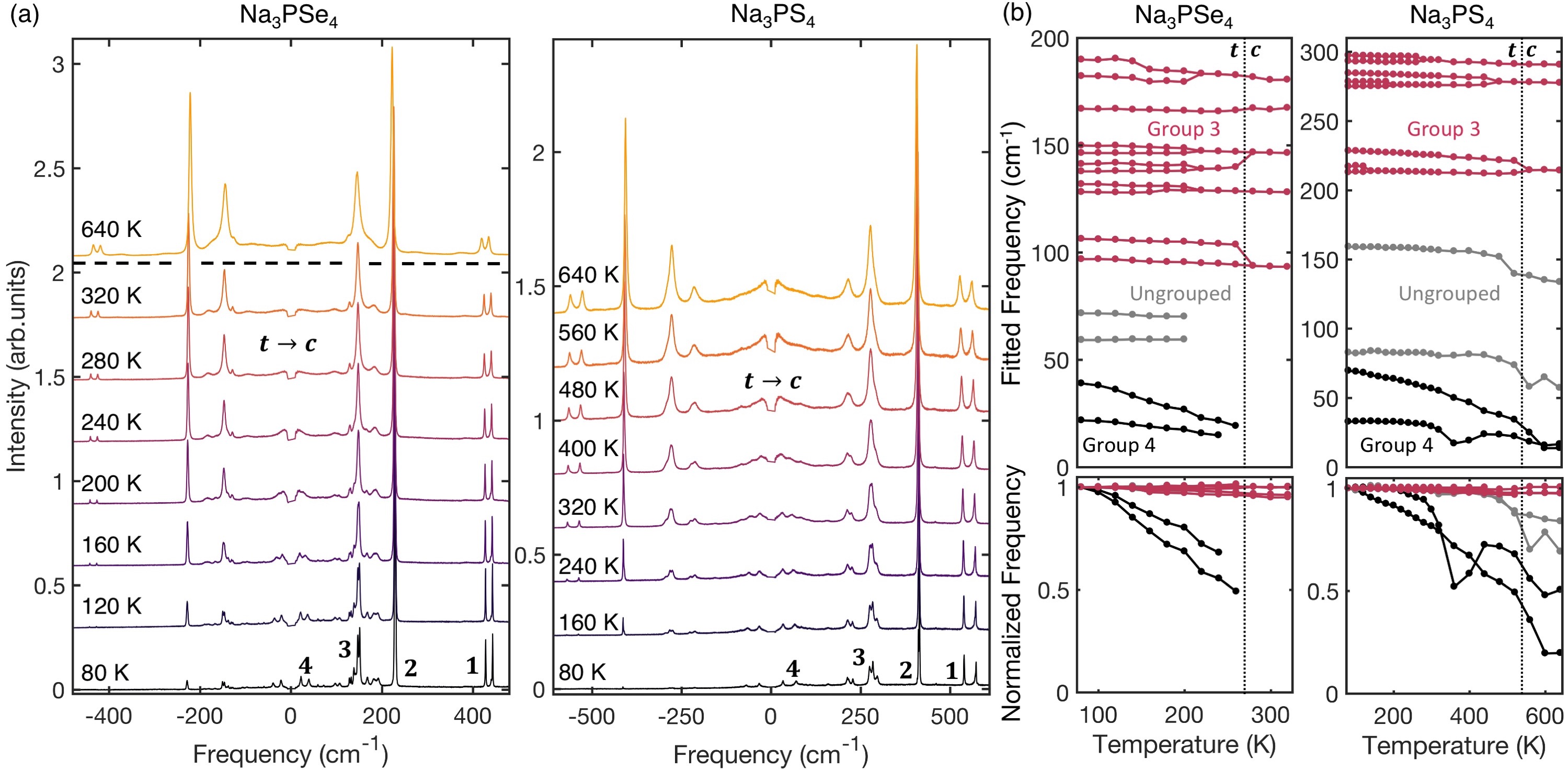}
     \caption{(a) Selected Raman spectra as a function of temperature for Na$_3$PSe$_4$ (between 80-320~K) and Na$_3$PS$_4$ (80-640~K), covering the \textit{t-c} phase transition in both cases (see \textit{t-c} label). The bold numbers 1 – 4 mark groups of features shared by both materials.
     (b) Top panels: Fit-derived frequency as a function of temperature for the peaks in groups 3 (red) and 4 (black) for Na$_3$PSe$_4$ (left) and Na$_3$PS$_4$ (right).
     The \textit{t-c} transition is marked by a dashed line. Bottom panels: same as the top panels but showing the fractional change of frequency. The peaks in group 4 (black) show anomalously strong shifts in relative frequency compared to the other modes in both compounds. }
     \label{Fig:Raman_spectra_all}
 \end{figure*}

In this study, we investigate the temperature-dependent lattice dynamics of the foundational sodium ion conductors Na$_3$PS$_4$ and Na$_3$PSe$_4$ through THz-range Raman scattering and first-principles calculations.
In both compounds, we clearly identify anharmonic vibrational modes involving coupled host lattice-Na$^+$ ion motion that drive the tetragonal-to-cubic (\textit{t-c}) phase transition, as reported previously in Na$_3$PS$_4$~\cite{Gupta:2021a,Famprikis:2021a}. 
Moreover, we demonstrate regime-crossing tunability of the material’s anharmonic character through the conceptually simple homovalent substitution of S for Se: 
while the phase transition is displacive in Na$_3$PSe$_4$, Na$_3$PS$_4$ exhibits dynamic symmetry breaking in a phase that is cubic only on average.
Our findings demonstrate that the anharmonic lattice dynamics of SSICs can exhibit different underlying mechanisms even for seemingly minor substitutional changes.

\section*{Results and discussion}

Stoichiometric Na$_3$PS$_4$ and Na$_3$PSe$_4$ are known to take on tetragonal and cubic phases depending on temperature (see Fig.~\ref{Fig:Na3PCh4_Structure}), with Na$_3$PS$_4$ having recently been discovered to also possess a plastic polymorph~\cite{Krauskopf:2017aa,Pompe:2016thesis,Nishimura:2017a,Famprikis:2019a,Famprikis:2021a}.
The cubic structure (space group \hmn{I-43m}, $T_d^3$, \#217) is composed of PCh$_{4}^{3-}$ (Ch=chalcogen) tetrahedra arranged on a BCC lattice with P atoms at BCC lattice sites and Na$^+$ ions located at face centers and edges of the BCC cube.
In the tetragonal structure (space group \hmn{P-42_1c}, $D_{2d}^4$, \#114) the PCh$_{4}^{3-}$ tetrahedra are tilted about the crystallographic c-axis while a subset of the Na$^+$ ions are offset along the c-axis above and below their positions in the cubic phase.

Fig.~\ref{Fig:Raman_spectra_all}a shows the Raman spectra of Na$_3$PSe$_4$ and Na$_3$PS$_4$, normalized to the maximum intensity mode, throughout the temperature range encompassing both the tetragonal and cubic phases of each compound.
Both materials show a similar set of features numbered in bold for the lowest temperature measurement as follows: a pair of peaks at very high frequency (1), a sharp and intense single dominant peak (2), a group of intermediate frequency modes (3), and a pair of very low-frequency modes (4).
First-principles calculations based on density functional theory (DFT, see Methods) of the Raman spectra find a set of features similar to experiment for both compounds (see Fig. S1).

To quantify changes to the experimental spectra with temperature, we fit each spectrum with a multi-Lorentz oscillator fit (see Methods) in order to extract each peak’s temperature-dependent frequency.
It was found that groups 1 and 2 show very little change with temperature, whereas the peaks of group 3 (Fig.~\ref{Fig:Raman_spectra_all}b) gradually merge as temperature is increased toward the \textit{t-c} transition.
We defined the temperature of the \textit{t-c} transition ($T_c$) as occurring after the last peak merging.
This occurred between 260-280K and between 520-560K in Na$_3$PSe$_4$ and Na$_3$PS$_4$, respectively, both in agreement with X-ray diffraction measurements~\cite{Famprikis:2021a,Famprikis:2019a,Nishimura:2017a,Krauskopf:2017aa,Krauskopf:2018aa}.

\begin{figure*}
     \centering
     \includegraphics[width=0.65\textwidth]{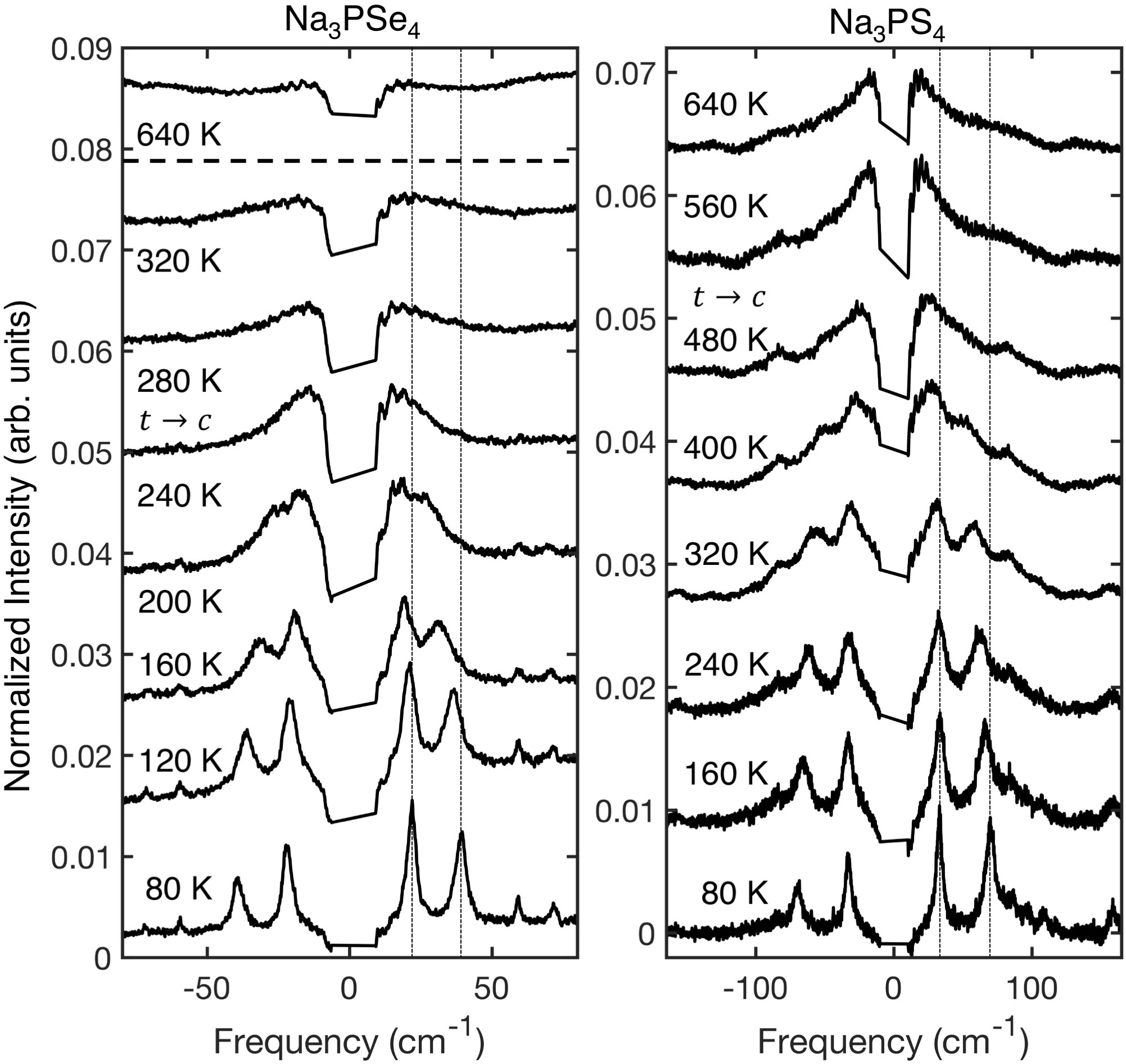}
     \caption{Low-frequency region of the Raman spectra as a function of temperature, covering the \textit{t-c} phase transition (indicated in figure) for  Na$_3$PSe$_4$ and  Na$_3$PS$_4$. This region contains two soft modes which shift toward 0~\wns as the temperature is raised. The two dashed vertical lines mark the location of the two soft modes at 80~K and serve as a guide to the eye for observing the shifts of each peak. The spectrum of Na$_3$PSe$_4$ at 640K is included for direct temperature comparison to Na$_3$PS$_4$. All spectra have been normalized by the integrated intensity of the peak in group 2 in order to enable comparison of relative intensity to this peak.}
     \label{Fig:low_frequency_Raman}
 \end{figure*}

The peaks of group 4 exhibit a notable temperature dependence (Fig.~\ref{Fig:Raman_spectra_all}b).
As temperature is raised, the frequency of these modes decreases much more quickly than for any of the other modes (bottom panels in Fig.~\ref{Fig:Raman_spectra_all}b),
approaching zero frequency as T$_c$ is approached.
We investigate the mechanisms underlying the evolution of these modes by inspection of the temperature dependence of the low-frequency region of the Raman spectra (see Fig.~\ref{Fig:low_frequency_Raman}).
All the spectra in this figure have been normalized by the integrated intensity of the group 2 peak to enable comparison of relative intensity to this peak. 
At 80K the peaks of group 4 appear sharp and well resolved. 
As temperature is raised, the peaks redshift and broaden, eventually merging near the transition so the two peaks can no longer be distinguished.
Above the \textit{t-c} transition the behavior of the two materials diverges.
In Na$_3$PSe$_4$, the group 4 feature decreases in relative intensity compared to the group 2 peak and becomes broad and flat. This behavior persists up to 640K, the highest temperature measured.
In Na$_3$PS$_4$, the group 4 feature merges into one peak whose intensity relative to the group 2 peak remains relatively constant. 
The differing behavior of Na$_3$PSe$_4$ and Na$_3$PS$_4$ above the phase transition is a sign their structural dynamics are fundamentally different in character, as discussed further below. 

The behavior of the group 4 peaks can be explained by considering that they are soft modes which drive a displacive phase transition.
Famprikis et al. and Gupta et al. also observed that the \textit{t-c} transition in Na$_3$PS$_4$ is soft-mode driven~\cite{Famprikis:2021a,Gupta:2021a} and we additionally observe here that Na$_3$PSe$_4$ displays the same phenomenon.
In a displacive phase transition, a gradual shift (displacement) of the atoms with temperature is driven by anharmonic interactions between the soft mode and other vibrations excited in the crystal at a given temperature.
This eventually leads to a discontinuous change of symmetry in the structure at the phase transition, when the atoms arrive at the positions and symmetry of the new crystal structure~\cite{dove2003structure,Dove:1997aa,fultz_2020,Blinc1974,Makarova1984, Nathan2012, Gorev1986,Cohen_2022aa}.
Simultaneously, the frequency of the soft mode reaches zero as the phase transition is approached from temperatures both above and below the transition.
At the transition temperature, the oscillatory motions of the atoms involved become a stationary distortion of the structure resulting in the new symmetry.
Because of the symmetry changes involved in the transition, the soft mode usually appears as a single mode in the high-symmetry phase and as a pair of modes due to broken degeneracy in the low-symmetry phase.
The group 4 peaks have the characteristics of the low-symmetry phase soft-mode pair.
That the full decay to zero frequency is not observed here is because of non-idealities in the real material system and instrument limits.

Above the phase transition temperature, information about the crystal symmetry combined with the Raman selection rules shows that the soft mode pair is expected to merge into a single-frequency triply-degenerate soft mode located at the Brillouin zone boundary, which is not Raman active. 
Indeed, the DFT-calculated phonon dispersion of the cubic phase of both materials shows a lattice instability of a triply-degenerate phonon at the Brillouin zone boundary (Fig.~S2), which is not accessible by Raman.
The expected disappearance of the soft modes is observed in our experiments in Na$_3$PSe$_4$ but not in Na$_3$PS$_4$ (see Fig.~\ref{Fig:low_frequency_Raman}) – again emphasizing their differing structural dynamical character.

\begin{figure}
     \centering
     \includegraphics[width=0.95\linewidth]{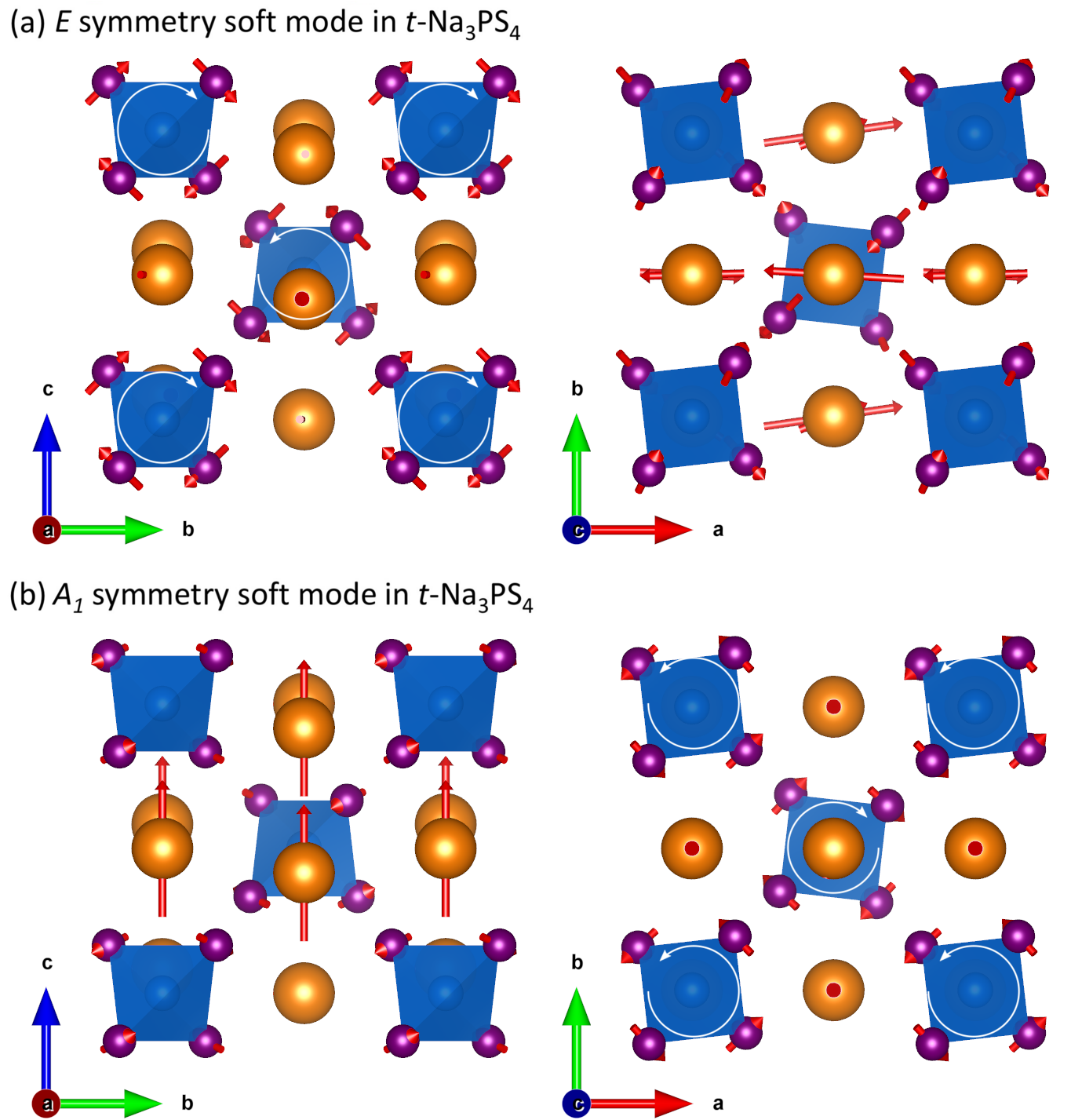}
     \caption{DFT-computed eigenvectors of the two soft modes identified from Raman spectroscopy in Na$_3$PS$_4$. In both cases the mode with wavevector $\mathbf{q}||c$ is shown, though other $\mathbf{q}$-directions show nearly identical eigenvectors. (a) The mode of $E$ symmetry corresponds to the lower frequency soft mode. This mode is doubly degenerate and only one of the two modes is shown. It involves rotation of the PS$_4^{3-}$ tetrahedra about the a-axis (left panel, white arrows) and Na$^+$ translation along the a-axis (right panel). The second degenerate mode has the same motion, but about the b-axis. (b) The mode of $A_1$ symmetry corresponds to the higher frequency soft mode. This mode involves rotation of the PS$_4^{3-}$ tetrahedra about the c-axis (right panel, white arrows) combined with Na$^+$ translation along the c-axis (left panel). The findings for the Se material are qualitatively similar (see Fig. S4).}
     \label{Fig:DFT_computed_modes}
 \end{figure}

Next, we extract the soft-mode eigenvectors to examine if it involves motion of the Na$^+$ mobile ion.
We identified the soft modes in our DFT-computed Raman spectra of the tetragonal phase (Fig.~S1) by their frequency and symmetry.
These modes are expected to be the two lowest frequency optical modes with single and double-degeneracy symmetries (in order to combine to triply degenerate).
Indeed, we find such a pair of modes in computational spectra of both materials (Table~S1).
In Fig.~\ref{Fig:DFT_computed_modes} we show the DFT-extracted eigenvectors of these modes for Na$_3$PS$_4$, with those of Na$_3$PSe$_4$ found in Fig.~S3. 
Interestingly, we find that there is coupling of mobile ion (Na$^+$) and host lattice dynamics since the soft mode eigenvectors in both compounds exhibit a combination of tetrahedral tilting and Na$^+$ translation with $A_1$ (high-frequency mode) and $E$ (low-frequency mode) vibrational symmetry (Table~S1).
For Na$_3$PS$_4$, this is in agreement with the results of Famprikis et al. and Gupta et al.~\cite{Famprikis:2021a,Gupta:2021a}.

The finding that collective tilting across the phase transition is mediated by the Na$^+$ ions also allows us to rationalize that a tilting-like transition occurs in Na$_3$PSe$_4$ and Na$_3$PS$_4$, despite their tetrahedra being isolated, which is different compared to the cases of corner- or edge-sharing structures (e.g. perovskites). 
With these findings, we have established coupling between the motion of the mobile ion and host lattice within the soft mode lattice instability.
Indeed, a number of prior works have discussed the role of coupled mobile ion-host lattice dynamics in ion conduction, including studies that have investigated  Na$_3$PS$_4$~\cite{Brenner:2020a,Zhang:2019aa,Zhang:2020aa,Adelstein:2016aa,Kweon:2017aa,Duchene:2019aa,Smith:2020aa,Jansen:1991aa,Lunden:1988aa,Gupta:2021a}.

Having identified  strong anharmonicity and connections to Na$^+$ dynamics in both compounds, we now compare the character of this anharmonicity between them.
Interestingly, while these compounds both appear to exhibit a displacive phase transition, a closer inspection of the low-frequency spectral range in Fig.~\ref{Fig:Raman_spectra_all}a, shown in Fig.~\ref{Fig:low_frequency_Raman}, indicates that Na$_3$PSe$_4$ and Na$_3$PS$_4$ display differing anharmonic character.
Above $T_c$, the soft modes disappear in Na$_3$PSe$_4$, leaving a flat, broad feature which remains unchanged up to 640~K and is attributed to second-order Raman scattering.
This is indeed what is expected for a purely displacive phase transition~\cite{dove2003structure}, since the cubic phase soft mode appears at the Brillouin zone boundary and therefore is Raman inactive (Fig.~S2).
However, in Na$_3$PS$_4$, the soft modes persist into the cubic phase, merging into a broad peak centered at zero frequency. 

The persistence of the soft mode feature into the cubic phase in Na$_3$PS$_4$, rather than its disappearance, indicates that the cubic structure observed in diffraction measurements is only a dynamically-averaged structure while the instantaneous, local structure exhibits lower symmetry~\cite{Gupta:2021a,Krauskopf:2018b}.  
Similar behavior has been observed in halide perovskites~\cite{Sharma:2020b,Yaffe:2017aa,Cohen_2022aa}, and PbMO$_3$ (M=Ti,Zr,Hf) perovskites~\cite{Roleder2000,Xu2019} where the appearance of first-order scattering in the cubic phase violates the Raman selection rules associated with the average structure.
Following our above finding that the soft mode shows Na$^+$-host-lattice vibrational coupling, we attribute the persistent soft mode feature in Na$_3$PS$_4$ to relaxational motion along this soft-mode eigenvector, in analogy to relaxational motion of the octahedral tilting modes found in halide perovskites~\cite{Sharma:2020b,Cohen_2022aa}.
This assessment is  supported by the strongly anharmonic thermal ellipsoids refined from synchrotron x-ray scattering experiments in the cubic phase~\cite{Nishimura:2017a} and by refinements of pair distribution function (PDF) measurements~\cite{Krauskopf:2018b,Famprikis:2021a}.
Thus, the cubic phase dynamically samples different tetrahedral tilting configurations.
The disorder resulting from this dynamic symmetry breaking causes a violation of the Raman selection rules for the soft mode.
In other words, for Na$_3$PS$_4$, our findings indicate a co-existence of displacive and order-disorder\footnote{`Order-disorder' is used here in the lattice dynamics sense~\cite{Dove:1997aa,Armstrong1989} rather than the alloy mixing sense~\cite{fultz_2020}.} phase transition characters~\cite{Dove:1997aa,Armstrong1989}.  

We note that we cannot confirm here any further selection rule violations in cubic Na$_3$PS$_4$ that might occur as a result of the lowered instantaneous symmetry, as previously proposed for the high frequency (550~\wn) modes~\cite{Famprikis:2021a}. The splitting of these modes can be explained by LO/TO splitting~\cite{YuPeter.Y;Cardona2005} that persists into the cubic phase (Figure S2).

This diverging anharmonic character of these two compounds is surprising.
The crystal chemistry of the two compounds is very similar and both structures show strong covalent bonding within the PS$_4^{3-}$/PSe$_4^{3-}$ tetrahedra and ionic bonding between the Na$^+$ ions and the tetrahedra.
We note that the Shannon ionic radii of Se$^{2-}$ is slightly larger than S$^{2-}$ (1.98~\AA \ and 1.84~\AA), so size effects may play a role.
However the differences arise, these very similar structures show markedly different lattice dynamical behaviors.
This suggests that tuning anharmonic effects in solids can be very subtle, with a simple homovalent substitution changing the lattice dynamics qualitatively.

Our findings indicate that coupled anharmonic motion of the mobile ion and host lattice is an important structural dynamical feature of this class of sodium ion conductors, which display two qualitatively different manifestations of this anharmonicity in Na$_3$PSe$_4$ \textit{versus} Na$_3$PS$_4$. 
Gupta et al. have shown that this particular anharmonic motion may assist ion conduction in this material through coordination of the jump process with dynamic modification of the host lattice bottleneck, arising from the nature of the soft mode motion~\cite{Gupta:2021a}.
This mechanism of conductivity enhancement does not require full rotary motion of the anions as in the paddlewheel effect and also affords a coordination of motion that the random spinning of anions does not\cite{Smith:2020aa,Zhang:2019aa,Zhang:2020aa,Famprikis:2019a}.
It is reasonable to surmise that the aliovalent doping that generates Na vacancies and record high conductivity in these compounds~\cite{Smith:2020aa,Zhang:2019aa,Zhang:2020aa,Famprikis:2019a} can also affect qualitative changes to the behavior of this lattice instability, and this is an area that requires further research.

\section*{Conclusions}

We combined THz-range Raman scattering and DFT calculations to compare the structural dynamics of the Na$^+$ ion conductors Na$_3$PS$_4$ and Na$_3$PSe$_4$, whose doped counterparts have recently demonstrated record Na$^+$ conductivities~\cite{Fuchs:2020a,Hayashi:2019vr}.
These compounds are isostructural and both compounds possess a tetragonal-to-cubic phase transition at elevated temperature.
Anharmonicities due to a vibrational instability in the cubic structure drive the phase transition.
In the tetragonal phase, both compounds show telltale soft mode behavior which indicates the instability of a single normal mode is the source of the phase transition.
Our computational findings show the soft modes involve the coupled motion of the mobile Na$^+$ ion and the host lattice, where the Na$^+$ ions mediate the tilting of the PCh$_{4}^{3-}$ tetrahedra.
Importantly, the structural dynamics of their cubic phases have divergent character.
Na$_3$PSe$_4$ shows almost purely displacive character with the soft modes disappearing in the cubic phase as the change of symmetry shifts these modes to the Raman-inactive Brillouin zone boundary.
Na$_3$PS$_4$ instead shows order-disorder character in the cubic phase, with the soft modes persisting through the phase transition and remaining active in Raman in the cubic phase, violating Raman selection rules for that phase. 
This indicates the cubic phase of Na$_3$PS$_4$ is only cubic on average and actually samples different atomic configurations in real time. 
While the origin of the diverging anharmonic behaviors is not yet clear, it is important to note that this substitution of a homovalent atom to form an isostructural material has led to dramatically different structural dynamics. 
The anharmonicity in this material and its tunability with substitution may both play an important role in the high conductivities of the doped compound and this is suggested as an important direction for further work. 

\section*{Methods}

\subsection*{Material Synthesis}
Na$_3$PS$_4$ and Na$_3$PSe$_4$ were synthesized by high temperature ampoule-synthesis.
All synthesis preparations were carried out in Ar-filled glove box and ampoules were dried under dynamic vacuum at 800~$^\circ$C for 2 hours to remove all traces of water. 
The starting materials Na$_2$S (Sigma-Aldrich, 99.98\%) and P$_2$S$_5$ (Sigma-Aldrich, 99\%) for Na$_3$PS$_4$ and Na$_2$Se (self-synthesized~\cite{Krauskopf:2017aa}  with an adjusted heating ramp of 3~$^\circ$C/hr), P (99.995\% trace metal basis, ChemPur) and Se (99.5\% trace metal basis, Alfa Aesar) in the case of Na$_3$PSe$_4$ were ground together in an agate mortar respectively.
The homogenized mixtures were pressed into pellets, filled into quartz ampoules (12~mm inner diameter) and sealed under vacuum.
Reactions were performed in a tube furnace at 500~~$^\circ$C  for 20 hours with a heating ramp of 30~~$^\circ$C/hour.
The obtained pellets were ground into powders and stored in a glovebox for further use.
\subsection*{Raman Scattering}
We performed Raman measurements on a custom-built Raman system designed for low-frequency Raman and collection of both Stokes and anti-Stokes scattering by using two notch filters (Ondax).
We used a 785~nm diode laser (Toptica XTRA II) at powers of 5~mW (Na$_3$PS$_4$) and 2~mW (Na$_3$PSe$_4$) focused on the sample with a 50x NIR objective (Nikon Plan Apo NIR-C 50x/0.42).
Beam powers were chosen so that beam heating (as measured by the Stokes/anti-Stokes ratio) was undetectable above the noise level at 80~K, the lowest measured temperature.
Due to the extreme sensitivity of both materials to air, moisture, and local beam damage when performing Raman under vacuum the following steps were performed.
For low-temperature measurements (80-320~K), a powder sample of the material was pressed flat onto a glass cover slip and loaded into an inert atmosphere chamber inside a nitrogen-filled glove box.
The chamber consisted of a stainless steel blank bottom and optical window top that were sealed together with a KF-flange (copper gasket).
The KF-flange seal ensured that this chamber remained sealed even when placed under high vacuum conditions, keeping the sample in a gaseous atmosphere.
For Na$_3$PS$_4$, nitrogen from the glovebox was used as the working gas.
For Na$_3$PSe$_4$ the working gas used was helium.
This was achieved by first sealing the chamber with a rubber o-ring gasket, then transferring it to a glovebag where the atmosphere was exchanged for helium and then sealing with a copper gasket.
For high-temperature measurements ($>$373~K), Na$_3$PS$_4$ and Na$_3$PSe$_4$ were flame-sealed inside a glass capillary tube under argon.
This was done due to the spontaneous vaporization of some element(s) of Na$_3$PS$_4$ at temperatures above about 100~$^o$C (373~K).
The sealed capillary prevented release of any elemental vapors and enabled an equilibrium vapor concentration.

Raman measurements at low temperatures (80-320~K) were performed by mounting the inert atmosphere cell with sample inside onto the cold-finger of a cryostat (Janis, ST-500).
The cryostat was pumped to high vacuum before low-temperature measurements commenced.
Temperature was controlled with a Lakeshore temperature controller (model 335) with liquid nitrogen as the coolant.
The sample temperature was calibrated against the cryostat set temperature by measuring the temperature of the inert cell directly using a temperature gauge. 
The inert cell temperature was found to be <5C higher than the cryostat set point for all temperatures. 
Raman measurements at high temperatures (300-640~K) were performed by placing the Na$_3$PS$_4$ powder capillary onto the stage of a Linkam Temperature Controlled Stage (THMS600).
Temperature was controlled by the Link software.
Due to the slight risk of sulfur-containing vapors in case of a capillary burst during heating, precautions must be taken by having a well-ventilated room, the minimum amount of sample possible should be used, the capillary thickness should be suitable to the goal temperature, and the Linkam should be purged with inert gas to slow any reactions that may occur after breakage.
To ensure compatibility of the high-temperature and low-temperature data sets, measurements were overlapped at temperatures of 300~K and 320~K.

The peak widths, positions, and Stokes/anti-Stokes ratio were found to be in agreement for both methods.
The Raman spectra of both Na$_3$PS$_4$ and Na$_3$PSe$_4$ displayed a weak but noticeable background, which extended thousands of wavenumbers beyond the region of the Raman spectrum on the Stokes side.
This indicates the presence of fluorescence or phosphorescence, likely from defects.
This background was removed by specifying regions without Raman scattering, fitting a polynomial to these regions, and then subtracting the fitted polynomial.
These backgrounded spectra are displayed in the main text.
After background subtraction, the spectrum was fit with a multi-peak model.
Both the Stokes and anti-Stokes scattering were fit simultaneously in order to verify that all features indeed arise from Raman scattering and to verify that the temperature changes monotonically.
For Na$_3$PS$_4$, a damped Lorentz oscillator model was successful in fitting the peaks for all temperatures.
The damped Lorentz oscillator, rather than a Lorentzian, was required to capture the broad features at low wavenumber where the Lorentzian approximation doesn’t hold.
For Na$_3$PSe$_4$, we observed that the peaks could not be fit by either a pure damped Lorentz oscillator or a pure Gaussian peak shape. 
A pseudo-Voigt peak shape composed of a linear combination of a damped Lorentz oscillator and a Gaussian was employed to model the features of this sample.
This suggests the peaks of this sample are broadened by both lifetime and disorder-induced broadening.
The relative weight of the Lorentz oscillator increased with temperature, supporting this hypothesis.
The fitted equation for both materials can be expressed as:
\begin{multline}
    I_{Raman}(\omega,T) = \\ S_{BE}(\omega,T)\sum_j\Big((1-h_{j}) \frac{c_j|\omega|(\gamma_{L,j}^2/\omega_j)}{\omega^2\gamma_{L,j}^2+(\omega^2-\omega_j^2)^2}+\\h_{j}c_{j}exp[\frac{-4ln2(|\omega|-\omega_j)^2}{\gamma_{G,j}^2}]\Big)
    \label{Lorentz}
\end{multline}
where $I$ is the Raman intensity, $\omega$ is the Raman shift, and $T$ is the temperature. 
For the $j^{th}$ pseudo-Voigt peak, $\omega_{j}$ is the resonance frequency, $\gamma_{L,j}$ and $\gamma_{G,j}$ are the damping coefficients of the Lorentz oscillator and Gaussian components of the pseudo-Voigt, $c_j$ is the intensity coefficient, and $h_j$ is the relative weight of Lorentz oscillator versus Gaussian character. 
A value of $h_j=0$ gives pure Lorentz oscillator character and a value of $h_j=1$ gives pure Gaussian character. 
The sum of peaks is multiplied by the appropriate Bose-Einstein population factor ($S_{BE} (\omega,T)$) for Stokes and anti-Stokes scattering to account for the temperature dependence of the phonon populations. 
The lowest temperature spectrum was fit first, and then the fit to spectra at subsequent temperature steps was adapted from the previous step. When a pair of peaks could no longer be distinguished from a single peak, the fit to that feature was reduced to one peak.

Damped Lorentz oscillators oscillate at a frequency lower than their resonant frequency because of the damping. 
The fitted frequencies plotted in Fig.~\ref{Fig:Raman_spectra_all}c are the actual oscillation frequency ($\omega_{osc}$) which are corrected by the damping coefficient to be:
\begin{equation}
    \omega_{osc}=\sqrt{\omega^2-\gamma^2}
\end{equation}

\subsection*{First-principles calculations}
Calculations of zero-temperature phonon properties for both materials were performed using DFT.
We applied the projector-augmented-wave method~\cite{kresse_joubert_1999} as implemented in the VASP code~\cite{kresse_furthmuller_1996,kresse_furthmuller_1996a}, with exchange-correlation described by the Perdew-Burke-Ernzerhof (PBE) functional~\cite{perdew_etal_1996}.
In all calculations, the plane-wave energy cutoff was set to 350~eV, and the energy threshold for electronic convergence was set to $10^{-6}$~eV.

Both the unit-cell and the internal geometry of the two crystals were relaxed using the Gadget code \cite{bucko_etal_2005}, resulting in structures with forces smaller than  $\approx 10^{-3}$~eV/\AA.
A $\Gamma$-centered $11 \times 11 \times 11$ $k$-point grid was used in these relaxations.
Furthermore, we geometrically constrained the crystal lattice vectors to maintain a cubic or tetragonal symmetry, respectively.

Phonon frequencies and eigenvectors were obtained by a finite-displacement method using the phonopy suite~\cite{togo_tanaka_2015}, with a supercell size of 128 atoms.
To ensure the relatively tight settings that are required for phonon calculations, we used a $6 \times 6 \times 6$ grid for computing force constants and an $11 \times 11 \times 11$ grid for computing Born effective charges.
Non-analytic corrections based on dipole-dipole interaction~\cite{gonze_etal_1994,gonze_lee_1997} were included \textbf{}in order to correctly reproduce LO/TO splitting at the $\Gamma$ point in the tetragonal phases.

Phonon-based Raman spectra were computed by performing polarizability calculations~\cite{baroni_resta_1986,gajdos_etal_2006} for each Raman-active phonon mode, using the phonopy-spectroscopy tool \cite{skelton_etal_2017} (see SI).
For these calculations, the $k$-point grid was reduced to $4 \times 4 \times 4$, which we have verified to still guarantee sufficient numerical convergence.
In the tetragonal phases, the $\mathbf{q}$-direction dependence of the phonon modes near $\Gamma$ that arises due to LO/TO splitting was accounted for using a spherical integration procedure based on $7$th order Lebedev-Laikov quadrature~\cite{popov_etal_2020}, which allowed for obtaining spectra that are spherically averaged over $\mathbf{q}$.

\section*{\label{Acknowledgments}Acknowledgements}
O.Y. acknowledges funding from: ISF(209/21),Henry Chanoch
Krenter Institute, Soref New Scientists Start up Fund, Carolito Stiftung, Abraham \& Sonia Rochlin Foundation, E. A. Drake and R. Drake and the Perlman Family.
DAE acknowledges funding from: the Alexander von Humboldt Foundation within the framework of the Sofja Kovalevskaja Award, endowed by the German Federal Ministry of Education and Research; the Technical University of Munich - Institute for Advanced Study, funded by the German Excellence Initiative and the European Union Seventh Framework Programme under Grant Agreement No. 291763; the Deutsche Forschungsgemeinschaft (DFG, German Research Foundation) under Germany's Excellence Strategy - EXC 2089/1–390776260; Gauss Centre for Supercomputing e.V. for funding this project by providing computing time through the John von Neumann Institute for Computing on the GCS Supercomputer JUWELS at J\"ulich Supercomputing Centre. 
P.T. and W.G.Z acknowledge the Deutsche Forschungsgemeinschaft under grant ZE 1010/6-1

\bibliography{Bib_Sodium_IC}

\end{document}